\documentclass[aps,prl,amsmath,amssymb,reprint,superscriptaddress,preprintnumbers,showpacs]{revtex4-1}
\pdfoutput=1
\usepackage{bm,mathrsfs}
\usepackage[mathcal]{euscript}
\usepackage[pdftex]{hyperref}
\usepackage{dcolumn}
\usepackage{graphicx}
\usepackage{soul,color}
\setulcolor{red}
\setstcolor{blue}
\sethlcolor{green}
\renewcommand{\vec}[1]{\bm{#1}}
\begin{document}



\title{Current induced vortex superlattices in nanomagnets}
\author{Oleksii M. Volkov}
    \affiliation{Taras Shevchenko National University of Kiev, 01601 Kiev, Ukraine}

\author{Volodymyr P. Kravchuk}
 \email[Corresponding author. Electronic address: ]{vkravchuk@bitp.kiev.ua}
    \affiliation{Institute for Theoretical Physics, 03143 Kiev, Ukraine}

\author{Denis D. Sheka}
     \affiliation{Taras Shevchenko National University of Kiev, 01601 Kiev, Ukraine}
     \affiliation{Institute for Theoretical Physics, 03143 Kiev, Ukraine}

\author{Yuri Gaididei}
    \affiliation{Institute for Theoretical Physics, 03143 Kiev, Ukraine}

\date{\today}

%
%

\begin{abstract}
Influence of the spin-transfer torque on the vortex state magnetic nanodisk is studied numerically via Slonczewski-Berger mechanism. The existence of a critical current is determined for the case of same-directed electrical current, its spin polarization and polarity of the vortex. The critical current separates two regimes: (i) deformed but \emph{static} vortex state and (ii) essentially \emph{dynamic} state under which the spatio-temporal periodic structures can appear. The structure is a stable vortex-antivortex lattice. Symmetry of the lattice depends on the applied current value and for high currents (close to saturation) only square lattices are observed. General relations for sizes of the stable lattice is obtained analytically.
\end{abstract}




\maketitle


Vortices are ubiquitous in Nature. They play a fundamental role in various physical systems and one of their fascinating features is a possibility to create a variety of periodic structures. Examples are fluids \cite{Thomson1887}, superconductors of type II where vortices   and vortex lattices are created under the action of magnetic field \cite{Abrikosov04}; vortices and their arrays  nucleate in  rotating superfluid helium \cite{Donnelly91}  and   BEC \cite{Fetter09}.  Submicron size magnetic nanoparticles  give another  example where the vortices play crucial role. These nanoparticles  due to interplay between short-range exchange interaction and long-range  dipole interaction have a curling ground state \cite{Usov93}.  One way to control the behavior of nanomagnets is to pass a spin-polarized current through the nanodot. As was discovered by Slonczewski \cite{Slonczewski96} and Berger \cite{Berger96}, the spin-polarized current acts as an effective spin-torque, which is an important feature for design of electrically controlled devices of spintronics \cite{Zutic04,*Tserkovnyak05}.

 It was shown theoretically \cite{Caputo07,*Sheka07b,*Gaididei10,Ivanov07b,Choi10} and experimentally \cite{Choi10} that the dc spin-polarized current passing   perpendicularly to the  nanodisk can excite the circular motion of the vortex core. Such a circular motion can be excited \cite{Caputo07,*Sheka07b,*Gaididei10} under two conditions: the current density $|j|$ exceeds some critical value $j_c$ and $j\sigma p<0$, where $\sigma=\pm1$ is direction of spin polarization of the current (along the normal to the nanodisk $\hat{\vec{z}}$), and $p=\pm1$ is the vortex polarity (the direction of the vortex core magnetization). When  $j\sigma p>0$,  the vortex core remains at the center of the disk.

The aim of this Letter is to show that passing the spin-polarized electrical current with $j\sigma p>0$ through magnetic nanodisks can lead to creation of \emph{ periodic vortex-antivortex arrays}. This is a phenomenon with a threshold: it occurs when the current density exceeds some critical value which depends on material parameters of the nanomagnet and its sizes.
In the  Letter we report about numerical investigation of influence of the spin-polarized current on vortex state nanodisk  for wide range of current densities. Our micromagnetic simulations \footnote{We used the OOMMF code of version 1.2a4, \url{http://math.nist.gov/oommf/}. All simulations were performed for material parameters of permalloy: exchange constant $A=1.3\cdot10^{-11}$ J/m, saturation magnetization $M_s=8.6\cdot10^{5}$ A/m, the anisotropy was neglected and damping was chosen close to natural value $\alpha=0.01$ (with the exception of some specially mentioned in text cases with $\alpha=0$). The mesh cell was chosen to be $3\times3\times h$ nm. The current parameters $\sigma=+1$, $\eta=0.4$ and $\Lambda=2$ were the same for all simulations.} are based on the Landau-Lifshitz-Slonczewski equation \cite{Slonczewski96,Berger96,Slonczewski02}:
\begin{equation} \label{eq:LLS}
\dot{\vec{m}} = \vec m\times{\delta\mathcal{E}}/{\delta\vec{m}} + \alpha[\vec m\times\dot{\vec m}]-\sigma j \varepsilon \vec m\times[\vec m\times\hat{\vec z}].
\end{equation}
Here $\vec m = \left(\sin\theta\cos\phi,\sin\theta\sin\phi,\cos\theta\right)$ is a normalized magnetization, the overdot indicates derivative with respect to the rescaled time in units of $(4\pi\gamma M_s)^{-1}$, $\gamma$ is gyromagnetic ratio, $M_s$ is the saturation magnetization, and $\mathcal{E}=E/(4\pi M_s^2)$ is normalized magnetic energy. The normalized electrical current density $j=J/J_0$, where $J_0=M_s^2|e|h/\hbar$ with $e$ being electron charge, $h$ being the sample thickness and $\hbar$ being Planck constant. Here and below it is assumed that the current flow and its spin-polarization are directed along $\hat{\vec z}$-axis. The spin-transfer torque efficiency function $\varepsilon$ has the form $\varepsilon={\eta\Lambda^2}/{\left[(\Lambda^2+1)+(\Lambda^2-1)\sigma(\vec m\cdot\hat{\vec z})\right]}$, where $\eta$ is the degree of spin polarization and parameter $\Lambda$ describes the mismatch between spacer and ferromagnet resistance \cite{Slonczewski02,Sluka10}.

Studying response of the vortex state to the spin-transfer torque one can distinguish two critical current densities: $j_1$ and $j_2$. When $j<j_1$ the stationary state of the system  is a deformed vortex state (see below). When $j>j_2$ the system goes in a saturated state when all spins are aligned  along $\hat{\vec z}$ axis. When $j_1<j<j_2$ a rich variety of dynamic states is observed: the system demonstrates either chaotic dynamics of vortex-antivortex 2D gas or regular and stable vortex-antivortex lattices.

\begin{figure}
\includegraphics[width=0.8\columnwidth]{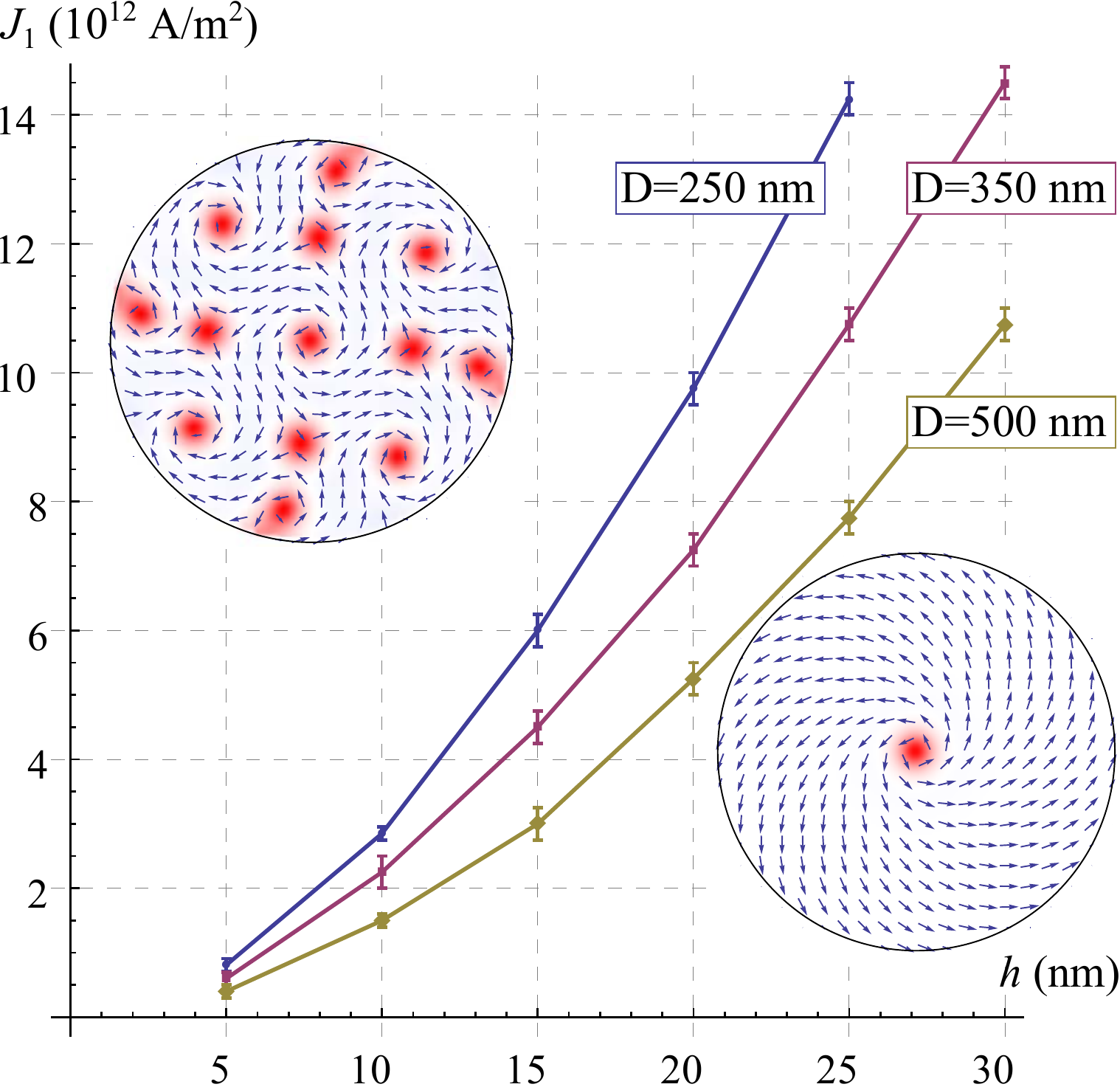}
\caption{(Color online.) Dependence of critical current $J_1$ on the nanodisk thickness for different disk diameters. Insets show magnetization structures in nanodisk with $D=250$~nm and $h=15$~nm for different current densities: $J=5.75\cdot10^{12}\;A/m^2\lesssim J_1$ (bottom right inset), $J=6\cdot10^{12}\;A/m^2\gtrsim J_1$ (upper left inset).}
\label{fig:jcrit}
\end{figure}

In the no-driving case the ground state of the nanodot is  a vortex state with  $\cos\theta_v=pf(r)$ where an exponentially localized function $f(r)$ describes the vortex core profile, and $\phi_v = \chi \pm {\pi}/{2}$ ( $(r,\,\chi)$ are polar coordinates  originated in the disk center). First of all we studied how the ground vortex state $\theta_v$, $\phi_v$ is modified under an influence of the  spin-polarized current. Starting from the current $j=0$ we increased $j$ step by step with full relaxation on each step and with size of the step $\Delta j\ll j_1$. A typical deformed vortex state is shown in the lower inset of the Fig.~\ref{fig:jcrit}. Changes of the out-of-plane component of the vortex distribution are negligibly small ($\theta\approx \theta_v$), while the in-plane structure is deformed appreciably and  can be described as $\phi=\phi_v+\psi(r)$. The deformation $\psi$ is a non-monotonic function  of $r$. Its amplitude is an increasing function of the current strength $j$. The presence of the deformation function $\psi(r)$ means an appearance of  volume magnetostatic charges and increase of corresponding magnetostatic energy. As a result at $j=j_1$, the vortex at the center of disk ceases to be a stable stationary state. The vortex escapes from the disk center moving in a spiral trajectory. The critical current $j_1$ is an increasing function of the disk thickness ( see Fig.~\ref{fig:jcrit}). This fact emphasizes  the important role of the volume magnetostatic charges in the phenomenon. The mentioned spiral vortex motion is accompanied by intensive generation of the magnons and results in appearance of a gas of vortex-antivortex pairs. As a rule the motion  vortices in this state is quite irregular. However, for some values of the current in the range $(j_1, j_2)$  instead of chaotic gas-like dynamics stable spatially regular vortex-antivortex structures   appear (as an example, see upper inset of the Fig.~\ref{fig:jcrit}). The following part of the Letter is devoted to studying this phenomenon.

To study the observed superstructures in detail we used two methods: (i) the spin current of the certain density is sharply applied to the vortex state nanodisk; (ii) the nanodisk is previously saturated to the uniform state by a strong external magnetic field which is applied perpendicularly to the disk plane, then the current of necessary density is switched on and the external field is adiabatically diminished down to zero: $\vec B=e^{-t/\tau}[B_0\hat{\vec{z}}+\vec{b}(\vec r,t)]$\footnote{$B_0=1.1$~T, $\tau=5$~ns, $\vec b=b_0\cos\omega t(\cos\chi,\,\sin\chi,\,0)$, where $b_0=0.01$~T and frequency $\omega=50$~GHz is much greater then typical eigenfrequencies of magnon modes in the disks of considered sizes.}, where $B_0$ is saturation field, time parameter $\tau$ is much greater than the typical time of spin waves dissipation and additional field $\vec b$ with $|b|\ll B_0$ was added to avoid instable equilibrium states.

\begin{figure*}
\includegraphics[width=\textwidth]{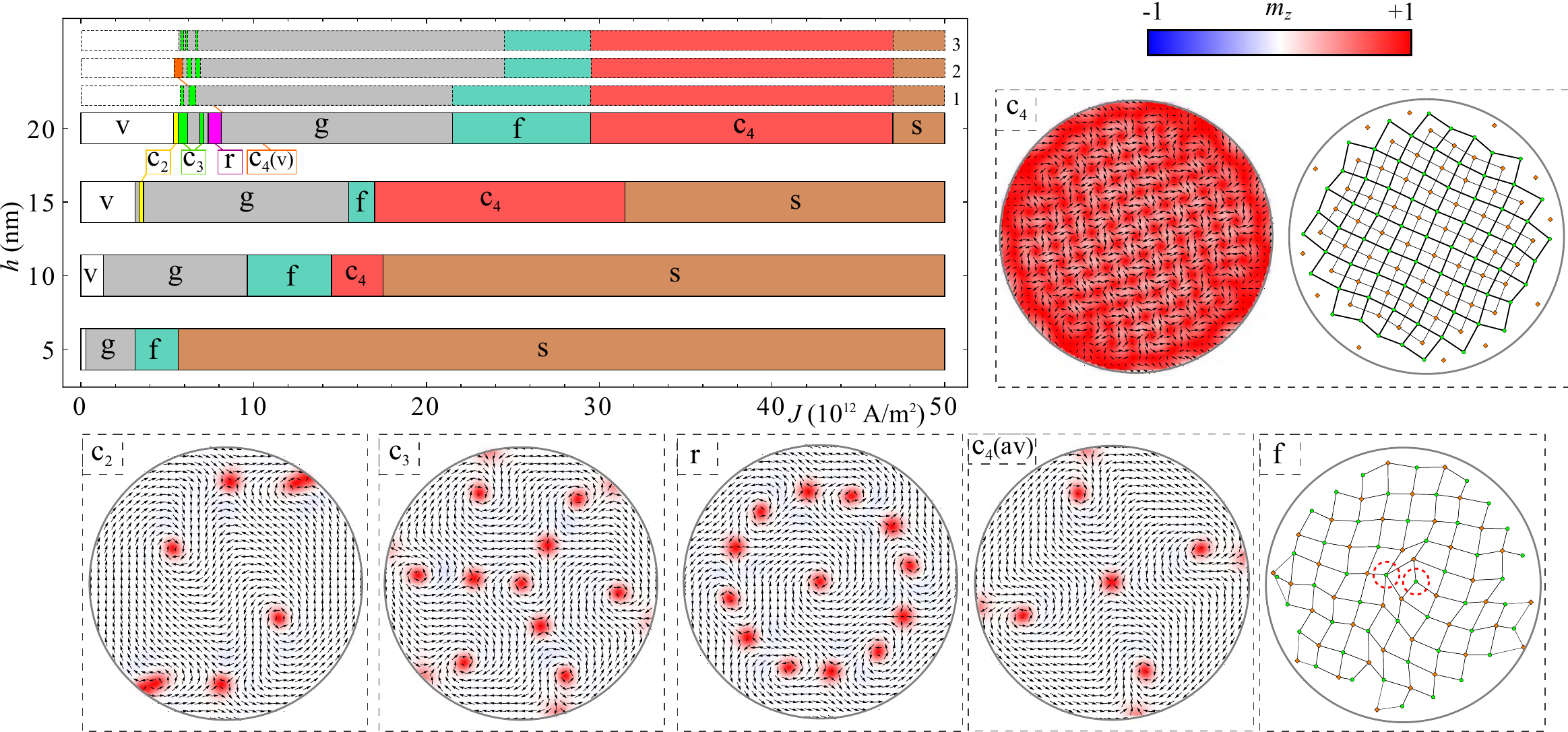}
\caption{(Color online.) The current diagrams of periodic structures in the nanodisk with diameter $D=350$~nm for thicknesses $h=5, 10, 15$ and 20~nm. The data are obtained using numerical simulations by the method (ii), see text. For $h=20$~nm three additional diagrams were obtained: 1 -- method (i) is used, 2 -- method (ii) with zero damping ($\alpha=0$), 3 -- method (i) with $\alpha=0$. The obtained states are classified in the following groups: ``v" -- static deformed vortex state, ``g"  -- regime of the chaotic dynamics of the vortex-antivortex gas, ``f" -- fluid-like dynamics (see text), ``s" -- saturation along the current direction. The other states are illustrated in the insets: arrows show the in-plane magnetization distribution and the out-of-plane component is shown by color (gray scale). Right part of the inset ``$c_4$" and inset ``f" present the vortex-antivortex lattices: rtices positions are shown by circles and antivortices -- by diamonds. All presented insets are obtained for thickness $h=20$~nm and for the following currents (in units $10^{12}\;A/m^2$): ``$c_2$" -- 5.5, ``$c_3$" -- 7, ``r" -- 8, ``$c_4$" -- 36, ``$c_4$(av)" -- 5.5 for $\alpha=0$, ``f" -- 24.}
\label{fig:diagram}
\end{figure*}

Possible vortex-antivortex structures which appear for $j_1<j<j_2$ are classified and shown in the Fig.~\ref{fig:diagram}. The obtained current diagram has the following properties: (i) the regular structures appear in the vicinity of the critical currents $j_1$ and $j_2$; (ii) the regions where the regular  structures exist, become wider with disk thickness increasing and these regions vanish for thin disks; (iii) regular structures have a well pronounced symmetry:  a system with rotary-reflection($S_2$) symmetry and  systems with rotational $C_n ~ (n=3,4,6)$ symmetry were observed. They  rotate bodily around the disk center with frequency much lower then the corresponding gyrofrequency of the free vortex. The magnitude of the rotation frequency and its sign depend on the current; (iv) the structures become more compact with the current increasing; (v) for the currents close to $j_2$ a crystal-like structure with a square lattice appears; (vi) as the current decreases line defects appear in such vortex-antivortex crystals, see inset ``f" of the Fig.~\ref{fig:diagram} (the defects are surrounded by dashed circles).
As a result the crystal loses the  long-range symmetry but a short-range order is preserved: each vortex has four nearest neighbors-antivortices and vice versa. The defects permanently move within the structure  and the whole dynamics is similar to the dynamics in fluids; (vii) further decrease of the current destroys the short-range order in the system and a vortex-antivortex gas phase appears.

It is important to note that the crystal-like structures appear in current driven nanodots even in the no-damping case ($\alpha=0$) and the interval of their existence only weakly depends on the damping coefficient $\alpha$ and does not depend on the method of structure creation.
In contrast to this, the thresholds and the intervals of existence of  regular structures with a few vortex-antivortex pairs  depend both on the damping coefficient and on the method of generation. These regular structures exist also in the no-damping case.

The total vorticity  $q=\sum_{k=1}^N\,q_k$, where $N$ is the total number of the particles: vortices (the winding number $q_k=+1$), antivortices ($q_k=-1$),  is the same for all regular structures and it is equal to $+1$ (as for the single vortex). The  structures with a few vortex-antivortex pairs also have $q=1$ if one takes into account  edge solitons. The edge soliton is a half of an antivortex which moves on the disk edge and it has the winding number $q_k=-1/2$.
  Each of above-mentioned regular vortex-antivortex structures represents a stationary state of the system. It means that the energy $\mathcal E$ and the total momentum \cite{Gaididei10b} $\mathcal J=\int(\cos\theta-1)(1-\partial_\chi\phi)\mathrm{d}\vec r$ must be time independent. Considering as the most intriguing  the no-damping case, one can obtain   from equations \eqref{eq:LLS} that
  \begin{eqnarray}\label{eq:E_J}
\dot{\mathcal{E}}=-\sigma j\int\,\varepsilon\,\sin^2\theta\,\dot\phi\mathrm{d}\vec r,\nonumber\\
\dot{\mathcal{J}}=\sigma j\int\,\varepsilon\,\sin^2\theta\big(1-\partial_{\chi}\phi\big)\mathrm{d}\vec r
\end{eqnarray}
 To describe the multiple vortex-antivortex structure the following ansatz can be used:
\begin{equation}\label{eq:ansatz}
\cos\theta=\sum_{k=1}^N\,f\big(|z-Z_k|\big),\quad \phi=\sum_{k=1}^N\,q_k\mathrm{arg}\big(z-Z_k\big),
\end{equation}
where  $z=re^{i\chi}$ and $Z_k=R_ke^{i\Phi_k}$ with $(R_k,\,\Phi_k)$ being polar coordinates of the $k$-th particle. Substituting \eqref{eq:ansatz} into the expressions \eqref{eq:E_J} and assuming that total area of the particles cores is much smaller than the disk area $\mathcal{S}$ one can obtain
\begin{eqnarray}\label{eq:E-J-a}
\dot{\mathcal{J}}\approx-\sigma j \xi\mathcal{J},
\quad \mathcal{J}\approx\pi h\sum_kq_kR_k^2+\mathcal{S}h(1-\sum_kq_k)\nonumber\\
\dot{\mathcal{E}}\approx-\pi h \sigma j \xi\sum_kq_kR_k^2\dot{\Phi}_k\;,
\end{eqnarray}
 where $\xi=\eta\Lambda^2/(\Lambda^2+1)$. Thus the stationarity conditions $\dot{\mathcal{E}}=0$ and $\dot{\mathcal{J}}=0$ can be satisfied if
\begin{equation}\label{eq:conditions}
q\equiv\sum_{k=1}^Nq_k=1,\quad\sum_kq_kR_k^2=0,\quad\dot\Phi_k=\Omega=\mathrm{const}.
\end{equation}
Moreover, in process of the structure formation $\mathcal{J}=\mathcal{J}_0\exp(-t/t_0)$, where $t_0=(\sigma\xi j)^{-1}$ is typical time of the formation, and systems with total vorticity $q=1$ have linear dependence between energy and momentum: $\mathcal{E}=\Omega\mathcal{J}$. For all regular structures obtained in the simulations $\sum_kq_k=1$ (taking into account edge solitons), $\Phi_k=\mathrm{const}$ and $\left|\sum_kq_k(R_k/L)^2\right|\lesssim10^{-2}$ with $L$ being the disk radius.
\begin{figure}
\includegraphics[width=0.8\columnwidth]{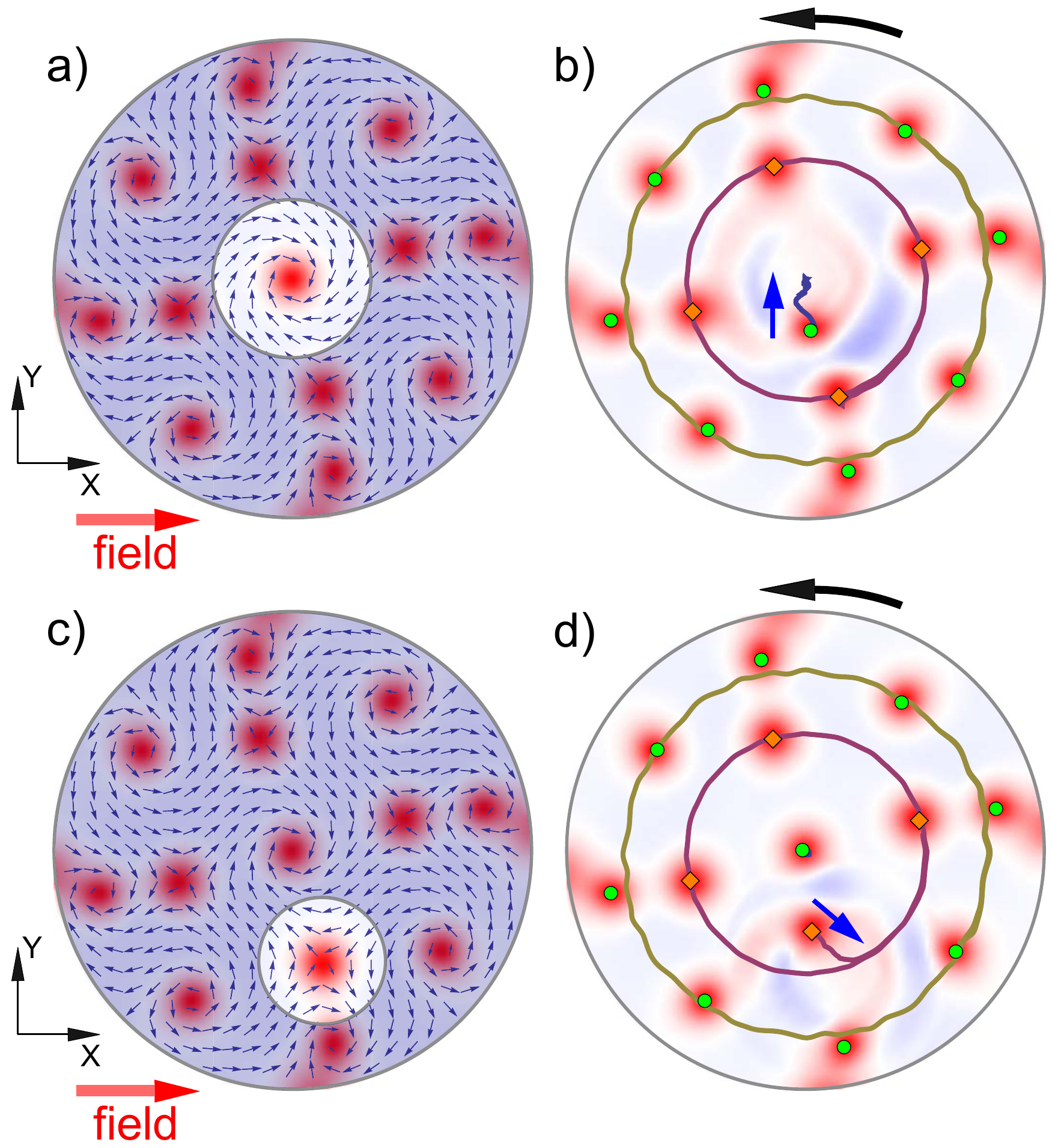}
\caption{(Color online.) Stability of the periodic structure in nanodisk with $D=250$~nm and $h=20$~nm.
Upper row illustrates the stability of the central vortex, and lower row --- the antivortex. Left column: the fixed (dark) and free (bright) magnetization distribution. The field pulse (red arrow) with the amplitude increasing up to $150$~mT was applied inside a free area during $t=0.75$~ns in order to shift a) vortex or c) antivortex. Right column: trajectories of the shifted vortex (antivortex) and its neighbors. The black arrow indicates the rotating direction of the structure. }
\label{fig:stability}
\end{figure}

 To validate our analytical considerations we carried out a direct stability check. To this end we first applied the field pulse in disk plane (red arrow), which was spatially localized nearby the central vortex in a circular range of 50 nm in diameter, keeping the rest of the magnetization distribution  artificially fixed during the field pulse time, see Fig.~\ref{fig:stability}a). Such a field pulse shifts a vortex perpendicular to the field direction. After the pulse was switched off the spin--current induced magnetization dynamics was simulated as usual. We found that the vortex was turned back to its origin in a short time (about 60ps) and a whole structure continued its rotation, see Fig.~\ref{fig:stability}b). The similar picture takes place for the antivortex stability, cf. Figs.~\ref{fig:stability}c) and Fig.~\ref{fig:stability}d). It allows to conclude that the superstructure is locally stable: only the strong field pulse can destroy the structure.

In conclusion, we observed for the first time and studied numerically the spatially periodic vortex--antivortex structures which  appear in nanomagnets under influence of spin-transfer torque. The structure formation is only weekly  influenced by Gilbert damping \footnote{The formation of vortex--antivortex patterns was confirmed also by spin--lattice simulations with nominal damping constant and without damping term.}, it results from effective dissipation induced by the spin current. We demonstrated analytically and proved numerically fulfillment of conditions \eqref{eq:conditions} for the periodic structures.

Authors acknowledge computing cluster of Kyiv National Taras Shevchenko University\footnote{Kyiv National Taras Shevchenko University high-performance computing cluster \url{http://cluster.univ.kiev.ua/eng/about}.} where the simulations were performed.

%

\end{document}